\documentclass[11pt]{article}
\usepackage[a4paper,margin=26mm]{geometry}
\usepackage[T1]{fontenc}
\usepackage{lmodern,amsmath,amssymb,amsthm,mathtools,mathrsfs}
\usepackage{microtype,booktabs,array,enumitem}
\usepackage{flafter,placeins}
\usepackage{xcolor,tikz}
\usetikzlibrary{arrows.meta,positioning,calc,decorations.pathreplacing,automata}
\usepackage[font=small,labelfont=bf]{caption}
\definecolor{inkblue}{RGB}{33,77,112}
\definecolor{lightblue}{RGB}{233,240,246}
\usepackage[colorlinks=true,allcolors=inkblue]{hyperref}
\hypersetup{pdftitle={Quadratic reset bounds for finite maximal prefix codes},pdfauthor={Samrath Singh Chadha}}
\newtheorem{theorem}{Theorem}[section]
\newtheorem{lemma}[theorem]{Lemma}
\newtheorem{proposition}[theorem]{Proposition}
\newtheorem{corollary}[theorem]{Corollary}
\theoremstyle{definition}

\theoremstyle{remark}

\newtheorem{example}[theorem]{Example}
\numberwithin{equation}{section}
\newcommand{\eps}{\varepsilon}
\newcommand{\Sig}{\Sigma}
\newcommand{\D}{\mathscr D}
\newcommand{\one}{\mathbf1}
\newcommand{\Z}{\mathbb Z}
\newcommand{\Q}{\mathbb Q}
\DeclareMathOperator{\Span}{span}
\DeclareMathOperator{\ord}{ord}
\DeclareMathOperator{\hull}{hull}
\setlist{topsep=3pt,itemsep=2pt}
\tikzset{>=Latex,every node/.style={font=\small},vertex/.style={circle,draw,minimum size=7mm,inner sep=1pt},marked/.style={vertex,draw=inkblue,fill=lightblue},flow/.style={->,semithick,draw=inkblue}}
\title{Quadratic reset bounds for finite maximal prefix codes}
\author{Samrath Singh Chadha}
\date{4 November 2026}
\begin{document}
\maketitle
\begin{abstract}
Let $C$ be a finite maximal prefix code of maximum codeword length $k\ge2$. We prove that, whenever the literal decoder of $C$ is synchronizing, it has a reset word of length at most $(k-1)(4k-5)$. The bound is independent of the alphabet and the number of decoder states. Its main ingredient is a short-detection theorem: a linear observation of the decoder, with terminal weight constant on states of depth $k-1$, is either identically zero or nonzero on a word of length at most $2k-3$. The proof uses a first-return decomposition in noncommuting variables and cancellation of a highest-degree coefficient.

For codes whose codeword lengths are $k-1$ and $k$, we prove the sharper bound \mbox{$2(k-1)^2-1$} and characterize equality. The binary code with the single short word $a^{k-1}$ attains this bound for every $k\ge2$; an age representation and a potential on circular arcs give the matching lower bound. Both upper bounds yield polynomial-time constructions in the explicit code size. The mathematical results and the algorithms are formalized in Lean, with polynomial work bounds in a specified binary cost model.
\end{abstract}
\section{Introduction}

A prefix code can be decoded as soon as a complete codeword has been read. If the starting position is unknown, however, different parsings may persist. A reset word removes this uncertainty: after reading it, every possible state of the decoder is the same. For the code $\{a,ba,bb\}$, the letter $a$ is a reset word. The decoder is either at the root or has just read $b$, and in both cases $a$ completes a codeword. For the code $\{aa,ab,ba,bb\}$, no word resets the decoder, since the two possible depth parities remain different.

We ask how long a shortest reset word can be when all codewords have length at most $k$. This parameter measures the depth of the prefix tree. It can be much smaller than the number of states: a full tree on $d$ letters can have $1+d+\cdots+d^{k-1}$ internal vertices. Thus even a polynomial bound in the number of states need not give a polynomial bound in $k$.

Throughout, $\Sig$ is a finite alphabet with at least two letters, and $C\subseteq\Sig^+$ is a finite nonempty maximal prefix code. Its \emph{literal decoder} has as states the proper prefixes of the codewords, including $\eps$. A transition extends the current prefix by one letter, except that completing a codeword returns the decoder to $\eps$. Maximality makes every transition defined. Write
\[
 k=\max_{c\in C}|c|,
 \qquad s(C)=\min\{|w|:w\text{ resets the literal decoder of }C\},
\]
where the second quantity is used only when a reset word exists.

\begin{theorem}\label{thm:universal}
If $k\ge2$ and the literal decoder of $C$ is synchronizing, then
\begin{equation}\label{eq:universal}
 s(C)\le(k-1)(4k-5).
\end{equation}
A reset word satisfying this bound can be constructed in time polynomial in the explicitly listed code and alphabet.
\end{theorem}

The next result treats all full prefix trees with leaves at two consecutive depths. For $r\ge1$ and $\varnothing\ne A\subsetneq\Sig^r$, set
\begin{equation}\label{eq:CA}
 C(A)=A\cup\bigl(\Sig^{r+1}\setminus A\Sig\bigr).
\end{equation}
Thus each word at depth $r$ is either a leaf in $A$ or has all its children as leaves at depth $r+1$.

\begin{theorem}\label{thm:two}
The code $C(A)$ is maximal, has height $r+1$, and is synchronizing. Moreover,
\begin{equation}\label{eq:two}
 s(C(A))\le2r^2-1,
\end{equation}
with equality if and only if $A=T^r$ for some nonempty proper subalphabet $T\subsetneq\Sig$. A reset word satisfying~\eqref{eq:two} can be constructed in polynomial time in the explicit input size.
\end{theorem}

The equality cases reduce, by identifying letters, to the following binary family.

\begin{theorem}\label{thm:lower}
For every $k\ge2$, the binary code
\[
 C_k=\{a^{k-1}\}\cup
 \bigl(\{a,b\}^{k}\setminus a^{k-1}\{a,b\}\bigr)
\]
has reset threshold $s(C_k)=2(k-1)^2-1$. The word
\[
 a^{k-1}\bigl(b^k a^{k-1}\bigr)^{k-2}
\]
is a shortest reset word. Inverse images under surjective letter maps give the same threshold over every finite alphabet with at least two letters.
\end{theorem}

Let $s_d(k)$ denote the largest reset threshold among synchronizing finite maximal prefix codes of height $k$ on a fixed $d$-letter alphabet. There are only finitely many such codes, and Theorem~\ref{thm:lower} ensures that the class is nonempty.

\begin{corollary}\label{cor:growth}
For every $d\ge2$ and $k\ge2$,
\[
 2(k-1)^2-1\le s_d(k)\le(k-1)(4k-5).
\]
In particular, $s_d(k)=\Theta(k^2)$.
\end{corollary}

\paragraph{Earlier work.}
Synchronization of prefix codes is part of the classical theory of codes; Sch\"utzenberger's work~\cite{Schutzenberger} studies synchronizing properties of a particular family of prefix codes. The shortest-reset-word problem for finite automata goes back to \v Cern\'y~\cite{Cerny}. Pin~\cite{Pin} develops the approach of bounding the length needed to reduce the image of a subset. The complementary approach enlarges preimages. Rystsov~\cite{Rystsov} uses this construction with linear algebra and averaging for regular automata; Steinberg~\cite{Steinberg} gives a general averaging lemma. Our universal bound combines preimage expansion with a word-length estimate specific to prefix trees.

Carpi and D'Alessandro~\cite{CarpiDA} explicitly study synchronization in terms of maximum word length. They relate bounds for synchronizing pairs of complete finite codes to bounds for incompletable words of finite languages. Their result is a conditional route to a quadratic bound in this parameter.

State bounds for prefix-code decoders also follow from results on one-cluster automata: automata in which repeated use of one letter leads every state to the same cycle. B\'eal, Berlinkov, and Perrin~\cite{BealBerlinkovPerrin} prove a quadratic state bound for this class and apply it to finite prefix codes. Berlinkov, Ferens, Ryzhikov, and Szyku\l a~\cite{Berlinkov} obtain an $O(N\log^3N)$ reset bound for literal automata of finite prefix codes, where $N$ is the number of states. Theorem~\ref{thm:universal} gives a bound directly in the height of a complete literal decoder.

The distinction between the automata matters. Here a reset word must merge every proper-prefix state, without minimizing the decoder. The related theorem on uncompletable words and matrix mortality~\cite{ChandelkarChadha} concerns incomplete codes. A finite maximal prefix code has no uncompletable word: every input can be extended on the right to a concatenation of codewords. Its reset problem therefore requires a different argument.

\paragraph{The proofs.}
For the general bound, repeated use of one letter first sends the decoder into a cycle of at most $k$ states. We then construct a reset word backwards, enlarging the preimage of a chosen state. Cyclic averaging supplies an enlargement as soon as some short word changes a preimage's cardinality. Section~\ref{sec:universal} carries out this construction using the short-detection theorem proved in Section~\ref{sec:detection}. The latter replaces a bound depending on the dimension of a state space by one depending only on tree depth.

For consecutive depths, the next codeword boundary is a nondecreasing function of the current boundary. Section~\ref{sec:consecutive} uses this order to advance the least remaining offset between two parsings. A closure property of the short codewords describes the only obstruction to the simplest advance. Section~\ref{sec:extremal} proves the exact binary lower bound by assigning a potential to sets of possible ages. Section~\ref{sec:algorithms} gives the algorithms. Appendix~\ref{app:arcs} contains the arc calculations, and Appendix~\ref{app:lean} describes the formal verification.

\section{The literal decoder}\label{sec:decoder}

Words act on states on the right. Thus $q\cdot uv=(q\cdot u)\cdot v$. For a set $S$ of states, write $S\cdot w=\{q\cdot w:q\in S\}$. A word $w$ is a reset word if $|Q\cdot w|=1$.

\begin{figure}[htb]
\centering
\begin{tikzpicture}[x=1cm,y=1cm]
 \node[vertex] (root) at (0,1.4) {$\eps$};
 \node[draw,rectangle,fill=lightblue] (a) at (-1,0) {$a$};
 \node[vertex] (b) at (1,0) {$b$};
 \node[draw,rectangle,fill=lightblue] (ba) at (.25,-1.3) {$ba$};
 \node[draw,rectangle,fill=lightblue] (bb) at (1.75,-1.3) {$bb$};
 \draw[->] (root)--node[left]{$a$}(a);
 \draw[->] (root)--node[right]{$b$}(b);
 \draw[->] (b)--node[left]{$a$}(ba);
 \draw[->] (b)--node[right]{$b$}(bb);
 \draw[flow] (2.8,.1)--node[above,align=center]{identify leaves\\with the root}(5.2,.1);
 \node[vertex] (e) at (6.4,.1) {$\eps$};
 \node[vertex] (s) at (8.8,.1) {$b$};
 \path[->] (e) edge[loop above] node{$a$}(e)
   edge[bend left=24] node[above]{$b$}(s)
   (s) edge[bend left=24] node[below]{$a,b$}(e);
 \node at (.4,-2) {prefix tree};
 \node at (7.6,-2) {literal decoder};
\end{tikzpicture}
\caption{The code $\{a,ba,bb\}$. Rectangles are leaves. Each transition into a leaf becomes a return to $\eps$; the letter $a$ sends both decoder states to $\eps$.}
\label{fig:decoder}
\end{figure}
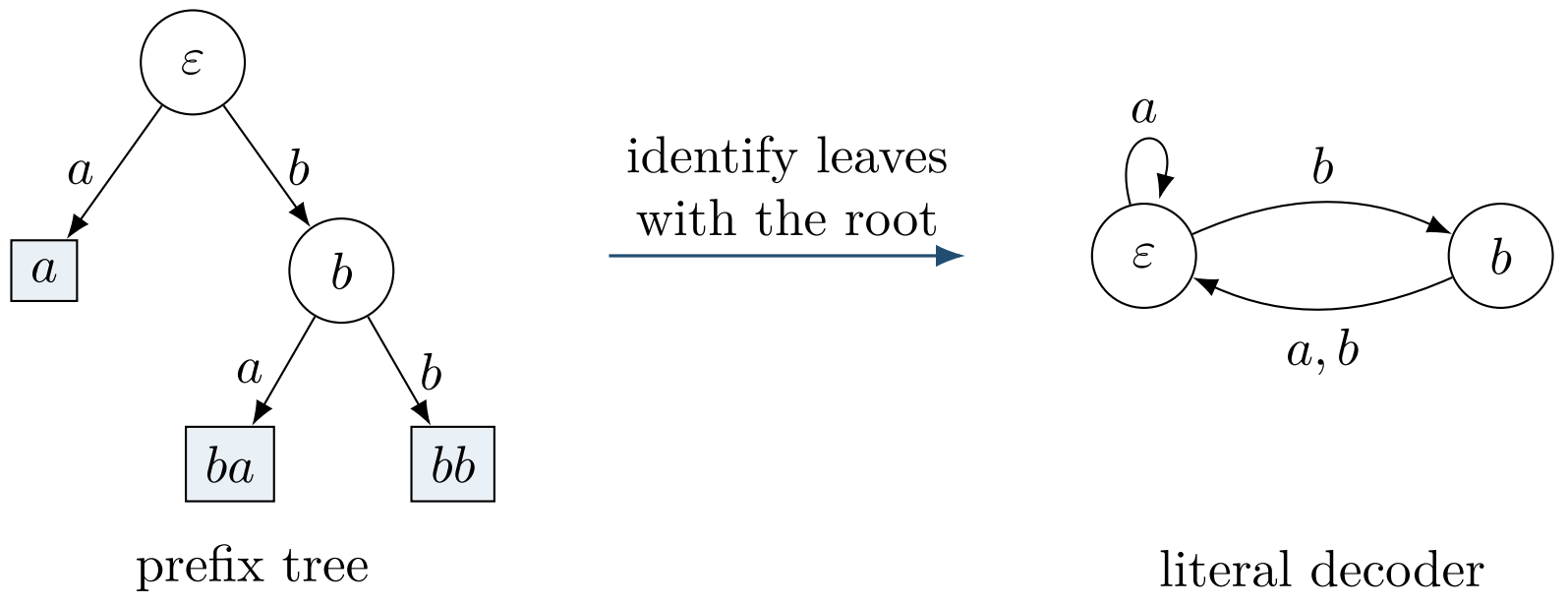

Let
\[
 Q=\{q\in\Sig^*:q\text{ is a proper prefix of a word of }C\}.
\]
The prefix tree of a finite nonempty prefix code is full if and only if the code is maximal. Indeed, a missing child of an internal vertex could be added as a codeword. Conversely, in a full finite tree, every word either meets a leaf or ends at a vertex having a leaf below it, so it is comparable with a codeword. Consequently, for $q\in Q$ and $b\in\Sig$, exactly one of $qb\in Q$ and $qb\in C$ holds, and
\begin{equation}\label{eq:decoder}
 q\cdot b=\begin{cases}qb,&qb\in Q,\\\eps,&qb\in C.\end{cases}
\end{equation}
Figure~\ref{fig:decoder} shows the passage from the tree to the decoder.

\begin{lemma}\label{lem:suffix}
Every state has depth at most $k-1$. Starting from a state $q$, every input of length $k-|q|$ completes a codeword. If at least one completion occurs while reading $w$, the final state is the suffix of $w$ after the last completion and has length at most $|w|-1$.
\end{lemma}
\begin{proof}
A path of length $k-|q|$ from $q$ that avoids all leaves would end at an internal vertex of depth $k$, which is impossible. At the last completion the decoder returns to $\eps$. Every subsequent letter extends the current prefix, since otherwise a later completion would occur. Those letters are therefore exactly the final state.
\end{proof}

\begin{lemma}\label{lem:normal}
If $|u|=k-1$, every state in $Q\cdot u$ is a suffix of $u$, and $|Q\cdot u|\le k$. If $u$ begins with a codeword, then $|Q\cdot u|\le k-1$.
\end{lemma}
\begin{proof}
Every nonroot initial state completes a codeword during $u$, by Lemma~\ref{lem:suffix}. The root either does the same or ends at the state $u$. Thus all outputs are suffixes of $u$, one for each possible length in $\{0,\ldots,k-1\}$. If $u$ begins with a codeword, every initial state completes at least once, so all output lengths are at most $k-2$.
\end{proof}

Reading from the root gives a unique factorization of any input into complete codewords followed by a proper prefix. This follows directly from~\eqref{eq:decoder}: the returns to $\eps$ mark the codeword boundaries. In particular, extending the final prefix to a leaf extends the input to a word of $C^*$.

If all leaves have depth $k$, fullness gives $C=\Sig^k$. Every letter then increases depth by one modulo $k$, so states of different depths cannot merge. For $k\ge2$, a synchronizing code must therefore contain a shorter leaf. At height one, $C=\Sig$ and the decoder has only the state $\eps$; its reset threshold is zero.

\subsection{A cycle reached by one repeated letter}

Fix $a\in\Sig$. Following the $a$-path from the root meets a unique leaf $a^r$, where $1\le r\le k$. The states
\[
 Z=\{\eps,a,\ldots,a^{r-1}\}
\]
form a cycle under $a$. Identify $Z$ with $\Z/r\Z$, and let $\rho$ be the rotation $i\mapsto i+1$. Set $h_0=k-1$ and define
\begin{equation}\label{eq:compression}
 \pi(q)=q\cdot a^{h_0},\qquad
 T_w(z)=z\cdot wa^{h_0}\quad(z\in Z).
\end{equation}
The map $\pi$ takes $Q$ into $Z$. The root already lies on the cycle, and every other state completes a codeword within $h_0$ letters, after which all remaining letters are $a$'s.

For later use, record the composition rules
\begin{equation}\label{eq:composition}
 T_{ua^{h_0}v}=T_v\circ T_u,
 \qquad T_{wa^j}=\rho^j\circ T_w,
 \qquad T_\eps=\rho^{h_0}.
\end{equation}
The first follows by writing the input as $ua^{h_0}va^{h_0}$. For the second, interchange the adjacent powers $a^j$ and $a^{h_0}$ and use the action of $a$ on $Z$. In particular, $T_\eps$ is a permutation.

\section{A universal bound from preimage expansion}\label{sec:universal}

We first explain how the decoder's height enters the reset construction. The only algebraic input is the following theorem, whose proof occupies Section~\ref{sec:detection}.

\begin{theorem}[Short detection]\label{thm:short}
Let $K$ be a field and $k\ge2$. Suppose $\lambda,\nu:Q\to K$, with $\nu$ constant on the states of depth $k-1$. Put
\[
 f(w)=\sum_{q\in Q}\lambda(q)\nu(q\cdot w).
\]
If $f$ is not identically zero, then $f(w)\ne0$ for some $|w|\le2k-3$.
\end{theorem}

To use this statement, take the cycle $Z$ and the maps $T_w$ from~\eqref{eq:compression}. We need to enlarge the preimage of any proper nonempty subset $S$ of $Z$. It is enough first to find a word $w$ for which $|T_w^{-1}(S)|\ne|S|$. Call such a word \emph{unbalancing for $S$}.

\begin{lemma}\label{lem:unbalance}
Assume that the decoder is synchronizing and $r\ge2$. For every $\varnothing\ne S\subsetneq Z$, there is a word $w$ of length at most $2k-3$ such that
\[
 |T_w^{-1}(S)|\ne |S|.
\]
\end{lemma}
\begin{proof}
Work over $\Q$ and set
\[
 \lambda(q)=\one_Z(q),\qquad
 \nu(q)=r\one_S(\pi(q))-|S|.
\]
A state of depth $k-1$ returns to $\eps$ on its next letter. Thus every such state has the same image $\eps\cdot a^{k-2}$ under $\pi$, and $\nu$ satisfies the hypothesis of Theorem~\ref{thm:short}. Its observation is
\[
 f(w)=r\bigl(|T_w^{-1}(S)|-|S|\bigr).
\]
If $z$ resets the decoder, $T_z$ is constant. Its preimage of $S$ has size either $0$ or $r$, so $f(z)\ne0$. Short detection gives the required $w$.
\end{proof}

\begin{lemma}[Cyclic averaging]\label{lem:average}
For any map $T:Z\to Z$ and subset $S\subseteq Z$,
\begin{equation}\label{eq:average}
 \sum_{j=0}^{r-1}|T^{-1}(\rho^{-j}S)|=r|S|.
\end{equation}
If $|T^{-1}(S)|\ne|S|$, some $0\le j<r$ satisfies
$|(\rho^j\circ T)^{-1}(S)|>|S|$.
\end{lemma}
\begin{proof}
Each point $T(z)$ belongs to exactly $|S|$ of the $r$ rotations of $S$. Counting over $z\in Z$ proves~\eqref{eq:average}. The summands have average $|S|$. If one differs from that average, one must exceed it.
\end{proof}

This is the averaging principle in the form needed here~\cite{Steinberg}. In view of~\eqref{eq:composition}, the rotation is realized by appending at most $r-1$ letters $a$. The first step can avoid this cost: a map that changes a singleton's preimage size is not a permutation, and hence has some fiber of size at least two.

\begin{proposition}\label{prop:cycle}
Suppose that $r\ge2$ and every proper nonempty $S\subset Z$ is unbalanced by some word of length at most $L$. Then the decoder has a reset word of length at most
\begin{equation}\label{eq:cyclebound}
 (k-1)+(r-1)(L+k+r-3).
\end{equation}
\end{proposition}
\begin{proof}
Choose an unbalancing word $w_0$ for a singleton. Since $T_{w_0}$ is not a permutation, choose $z_0$ with at least two preimages, and put
\[
 S_0=\{z_0\},\quad u_0=w_0,\quad S_1=T_{u_0}^{-1}(S_0).
\]
While $S_i\ne Z$, choose $|w_i|\le L$ unbalancing $S_i$. Lemma~\ref{lem:average} gives $0\le j_i<r$ such that, with $u_i=w_i a^{j_i}$,
\[
 S_{i+1}=T_{u_i}^{-1}(S_i),\qquad |S_{i+1}|>|S_i|.
\]
These sets need not be nested; only their cardinalities increase. After $m\le r-1$ steps, $S_m=Z$.

The word
\begin{equation}\label{eq:reset}
 W=a^{k-1}u_{m-1}a^{k-1}u_{m-2}\cdots a^{k-1}u_0a^{k-1}
\end{equation}
resets the decoder. Its initial power of $a$ sends $Q$ into $Z$. The rest acts there as $T_{u_0}\circ\cdots\circ T_{u_{m-1}}$, which sends $S_m$ into $S_0$. Notice that the maps are applied in the reverse order to their construction.

There are $m+1$ copies of $a^{k-1}$. Also $|u_0|\le L$ and $|u_i|\le L+r-1$ for $i\ge1$. Hence
\begin{align*}
 |W|&\le(m+1)(k-1)+mL+(m-1)(r-1)\\
 &\le k-r+(r-1)(L+k+r-2)\\
 &=(k-1)+(r-1)(L+k+r-3).
\end{align*}
\end{proof}

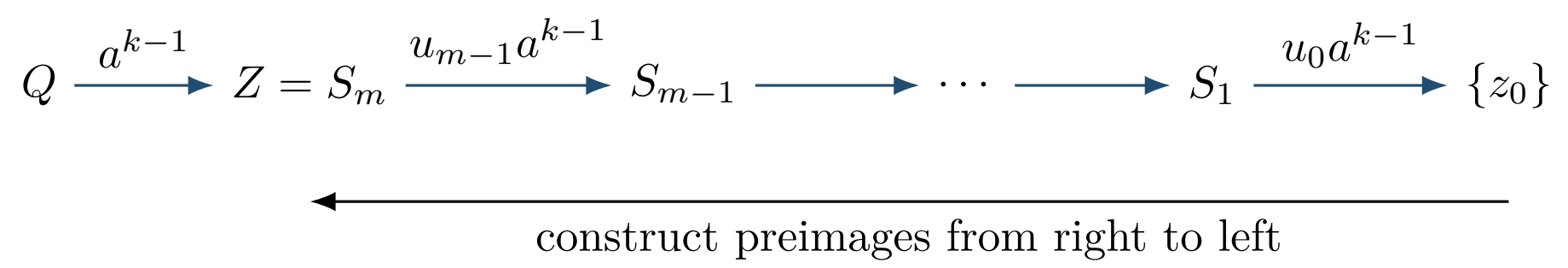
\begin{figure}[htb]
\centering
\begin{tikzpicture}
 \node (Q) at (0,0) {$Q$};
 \node (Z) at (2.1,0) {$Z=S_m$};
 \node (S) at (5,0) {$S_{m-1}$};
 \node (dots) at (7.2,0) {$\cdots$};
 \node (S1) at (9.1,0) {$S_1$};
 \node (S0) at (11.4,0) {$\{z_0\}$};
 \draw[flow] (Q)--node[above]{$a^{k-1}$}(Z);
 \draw[flow] (Z)--node[above]{$u_{m-1}a^{k-1}$}(S);
 \draw[flow] (S)--(dots);
 \draw[flow] (dots)--(S1);
 \draw[flow] (S1)--node[above]{$u_0a^{k-1}$}(S0);
 \draw[<-,semithick] (2.1,-.9)--node[below]{construct preimages from right to left}(11.4,-.9);
\end{tikzpicture}
\caption{The reset word reads from left to right. Its construction runs backwards: each $S_{i+1}$ is the full preimage of $S_i$ under $T_{u_i}$, and has larger cardinality.}
\label{fig:backwards}
\end{figure}

\begin{proof}[Proof of Theorem~\ref{thm:universal}]
If $r=1$, $a^{k-1}$ already resets $Q$. Otherwise Lemma~\ref{lem:unbalance} permits $L=2k-3$ in Proposition~\ref{prop:cycle}, giving
\[
 s(C)\le(k-1)+(r-1)(3k+r-6).
\]
For $2\le r\le k$, the product on the right increases with $r$: both factors are nonnegative and increasing. Substituting $r=k$ gives~\eqref{eq:universal}. The algorithmic assertion is proved in Section~\ref{sec:algorithms}.
\end{proof}

\begin{corollary}\label{cor:shortcycle}
If the decoder is synchronizing, $k\ge3$, and some pure-letter codeword $a^r$ satisfies $2r\le k+1$, then $s(C)\le2(k-1)^2-1$.
\end{corollary}
\begin{proof}
The case $r=1$ is immediate. Otherwise put $t=k-1$ and $u=r-1\le t/2$. The preceding estimate gives
\[
 s(C)\le t+u(3t+u-2)\le\frac74t^2\le2t^2-1,
\]
where the last inequality uses $t\ge2$.
\end{proof}

\section{Short detection from the height of the tree}\label{sec:detection}

We now prove Theorem~\ref{thm:short}. A first-return decomposition expresses every decoder observation using four word polynomials of degrees controlled by $k$. The difficulty is that these polynomials do not commute. We retain word order by embedding the free algebra in a division ring, where a finite linear system supplies a common right denominator. One further cancellation saves a degree and gives $2k-3$.

\subsection{Keeping the order of letters}

Write $K\langle\Sig\rangle$ for the free associative algebra on $\Sig$. Its elements are finite $K$-linear combinations of words, and multiplication is concatenation. Its completion $K\langle\!\langle\Sig\rangle\!\rangle$ consists of all formal word series. Products are well-defined because a word has only finitely many factorizations into two words.

\begin{lemma}\label{lem:embedding}
There is a division ring $\D$, with $K$ central, and an injective $K$-algebra homomorphism $\iota:K\langle\Sig\rangle\to\D$. If $R_n$ is the homogeneous component of degree $n$ of a word series $R$, then
\[
 R\longmapsto\widehat R(t)=\sum_{n\ge0}\iota(R_n)t^n
\]
is an injective multiplicative map into $\D[[t]]$, where $t$ is central. Its order is the least length of a word with nonzero coefficient in $R$.
\end{lemma}
\begin{proof}
Let $E=K(x_{b,i}:b\in\Sig,\ i\in\Z)$ be a rational function field in commuting algebraically independent variables. The assignment $\tau(x_{b,i})=x_{b,i+1}$ defines an automorphism of $E$ fixing $K$. Form the skew Laurent-series ring
\[
 \D=E((z;\tau)),\qquad zf=\tau(f)z\quad(f\in E).
\]
Its series have exponents bounded below. Each coefficient of a product is therefore a finite sum. Every nonzero series has an invertible lowest term $f_Nz^N$. Factoring out that term leaves $1+h$ with $h$ of positive order, whose inverse is the well-defined series $\sum_{j\ge0}(-h)^j$. Thus $\D$ is a division ring, and $K$ is central in it.

Send $b$ to $x_{b,0}z$. Then
\[
 \iota(b_1\cdots b_n)=x_{b_1,0}x_{b_2,1}\cdots x_{b_n,n-1}z^n.
\]
Distinct words of the same length give distinct monomials in the position-indexed variables. These monomials are linearly independent over $K$; different lengths have different powers of $z$. This proves injectivity. The additional variable $t$ records homogeneous degree. Multiplicativity follows by homogeneous convolution, and injectivity and the order assertion follow degree by degree.
\end{proof}

\begin{lemma}[Common right denominator]\label{lem:denominator}
Let $D,M\in\D[t]$, with $D\ne0$, $\deg D\le p$, and $\deg M\le m$. There are polynomials $U,V$ such that
\begin{equation}\label{eq:rightdenominator}
 DU=MV,\qquad V\ne0,\qquad\deg U\le m,\quad\deg V\le p.
\end{equation}
\end{lemma}
\begin{proof}
Write $U=\sum_{i=0}^m u_it^i$ and $V=\sum_{j=0}^p v_jt^j$. Equating coefficients in $DU-MV=0$ gives at most $m+p+1$ homogeneous equations in $m+p+2$ unknowns. This is a linear system over the right vector space $\D^{m+p+2}$: all fixed coefficients multiply the unknowns on the left. It has a nonzero solution. If $V=0$, then $DU=0$, forcing $U=0$, since the product of the leading coefficients of two nonzero polynomials is nonzero. Hence $V\ne0$.
\end{proof}

The side of multiplication in~\eqref{eq:rightdenominator} is essential. It gives $D^{-1}MV=U$ with precisely the order needed below.

\subsection{The first completion}

For each $q\in Q$, define
\[
 P_q=\sum_{s:\,qs\in C}s,\qquad
 M_q=\sum_{s:\,qs\in Q}\nu(qs)s,
 \qquad \mathcal F_q=\sum_{w\in\Sig^*}\nu(q\cdot w)w.
\]
The empty suffix contributes $\nu(q)$ to $M_q$. A word read from $q$ either completes no codeword or has a unique first completion. In the first case it contributes to $M_q$. In the second it consists of a suffix counted by $P_q$, followed by an input read from the root. Hence
\begin{equation}\label{eq:firstreturn}
 \mathcal F_q=M_q+P_q\mathcal F_\eps.
\end{equation}
Set $P=P_\eps$, $M=M_\eps$, and $D=1-P$. Since $P$ has no constant term, $D^{-1}=\sum_{j\ge0}P^j$ exists as a formal word series. Taking $q=\eps$ in~\eqref{eq:firstreturn} gives $\mathcal F_\eps=D^{-1}M$. Thus the observation series is
\begin{equation}\label{eq:renewal}
 F:=\sum_w f(w)w=A+BD^{-1}M,
\end{equation}
where
\[
 A=\sum_{q\ne\eps}\lambda(q)M_q,
 \qquad B=\lambda(\eps)+\sum_{q\ne\eps}\lambda(q)P_q.
\]
The tree gives the degree bounds
\begin{equation}\label{eq:degrees}
 \deg D\le k,\quad \deg M\le k-1,\quad
 \deg A\le k-2,\quad\deg B\le k-1.
\end{equation}
For example, a suffix in $M_q$ has length at most $k-1-|q|$, and $A$ uses only nonroot states. A suffix in $P_q$ has length at most $k-|q|$.

\subsection{Cancellation of the highest coefficient}

\begin{proof}[Proof of Theorem~\ref{thm:short}]
Use Lemma~\ref{lem:embedding} to encode the polynomials and series by a central variable $t$, retaining their names. Lemma~\ref{lem:denominator} gives
\[
 DU=MV,\qquad V\ne0,\qquad \deg U\le k-1,\quad\deg V\le k.
\]
The constant coefficient of $D$ is $1$, so it is invertible in $\D[[t]]$. Multiplying~\eqref{eq:renewal} on the right by $V$ gives
\begin{equation}\label{eq:numerator}
 FV=AV+BU=:N.
\end{equation}
By~\eqref{eq:degrees}, $\deg N\le2k-2$. We show that its coefficient of degree $2k-2$ vanishes.

Let $\omega$ be the common value of $\nu$ at depth $k-1$. Before encoding, put
\[
 \sigma=\sum_{b\in\Sig}b,\qquad
 J=\sum_{\substack{q\in Q\\|q|=k-1}}q,\qquad
 H=\sum_{\substack{q\in Q\\|q|=1}}\lambda(q)
       \sum_{\substack{qs\in Q\\|s|=k-2}}s.
\]
There is a leaf at depth $k$, so $J\ne0$. Every child of a state counted by $J$ is a leaf. Consequently the highest homogeneous components satisfy
\begin{equation}\label{eq:tops}
 D_k=-J\sigma,\qquad M_{k-1}=\omega J,
 \qquad A_{k-2}=\omega H,\qquad B_{k-1}=H\sigma.
\end{equation}
The last two equalities use the same paths: a nonroot initial state can contribute at those degrees only if it has depth one. The relevant path ends at a deepest internal vertex for $A$, and takes one additional letter for $B$. This also covers $k=2$, when the path contributing to $H$ is empty.

Now read~\eqref{eq:tops} as equalities in $\D$. The coefficient of degree $2k-1$ in $DU=MV$ is
\[
 -J\sigma U_{k-1}=\omega JV_k.
\]
Since $\omega$ is central, cancelling $J$ on the left yields
\[
 \sigma U_{k-1}+\omega V_k=0.
\]
Therefore the coefficient of degree $2k-2$ in $N$ is
\[
 \omega HV_k+H\sigma U_{k-1}
   =H(\omega V_k+\sigma U_{k-1})=0.
\]
We have proved $\deg N\le2k-3$.

If $F\ne0$, then $N=FV\ne0$. Orders of nonzero power series over a division ring add, because the product of their first nonzero coefficients is nonzero. Hence
\[
 \ord_t F+\ord_t V=\ord_t N\le\deg N\le2k-3.
\]
As $\ord_t V\ge0$, the first nonzero homogeneous part of $F$ has degree at most $2k-3$. Injectivity of the encoding gives an actual word of that length with $f(w)\ne0$.
\end{proof}

The terminal-depth condition is used exactly in~\eqref{eq:tops}. It makes both highest terminal coefficients scalar multiples of the corresponding path sums, allowing the cancellation. The conclusion concerns a linear observation; the extra compression and rotation costs in Section~\ref{sec:universal} are needed to turn it into a reset word.

\section{Codes with two consecutive lengths}\label{sec:consecutive}

Fix $r\ge1$ and $\varnothing\ne A\subsetneq\Sig^r$. The tree of $C(A)$ in~\eqref{eq:CA} is full: every word of length $r$ either belongs to $A$ or has all its children as leaves. Nonemptiness and properness ensure that both lengths $r$ and $r+1$ occur.

The direct construction in this section uses a property special to these two lengths. Along a fixed input, later codeword boundaries cannot have earlier next boundaries. We first describe this order and then show how to advance all surviving offsets between two parsings.

\subsection{Order of parsing boundaries}

Consider an infinite input $x_0x_1\cdots$. At a possible boundary $j$, set
\[
 e_j=\one_A(x_j\cdots x_{j+r-1}).
\]
The codeword starting there has length $r+1-e_j$, so its next boundary is
\begin{equation}\label{eq:boundary}
 F(j)=j+r+1-e_j.
\end{equation}
Since
\[
 F(j+1)-F(j)=1+e_j-e_{j+1}\in\{0,1,2\},
\]
the map $F$ and all its iterates are nondecreasing.

Let $p_0,p_1,\ldots$ be the boundaries of a reference parsing, so $p_{i+1}=F(p_i)$. If $p_i<q<p_{i+1}$ is an alternative boundary, then
\[
 p_{i+1}\le F(q)\le p_{i+2}.
\]
Equality at either endpoint means that the two parsings meet and agree thereafter. Otherwise the alternative boundary remains inside the next reference word. If its offset is $j=q-p_i$, the current reference word has length $\ell$, and the alternative word has length $\lambda_j$, the next offset is
\begin{equation}\label{eq:offset}
 j'=j+\lambda_j-\ell.
\end{equation}
In particular, when the reference word is short, the offset stays fixed or increases by one.

\begin{figure}[htb]
\centering
\begin{tikzpicture}[x=1cm,y=1cm]
 \draw[->] (-.4,0)--(11.6,0);
 \foreach \x/\lab in {0/p_i,5/p_{i+1},11/p_{i+2}}
  {\draw (\x,-.12)--(\x,.12);\node[below] at (\x,-.13) {$\lab$};}
 \node[above] at (2.5,1.1) {current reference word};
 \node[above] at (8,1.1) {next reference word};
 \fill[inkblue] (1.5,0) circle(2pt);
 \fill[inkblue] (7,0) circle(2pt);
 \draw[flow] (1.5,.12) to[bend left=23] node[above]{$F$}(7,.12);
 \node[below,inkblue] at (1.5,-.13) {$q$};
 \node[below,inkblue] at (7,-.13) {$F(q)$};
 \draw[decorate,decoration={brace,mirror,amplitude=4pt}] (0,-.85)--node[below=5pt]{$j$}(1.5,-.85);
 \draw[decorate,decoration={brace,mirror,amplitude=4pt}] (5,-.85)--node[below=5pt]{$j'$}(7,-.85);
\end{tikzpicture}
\caption{For lengths $r$ and $r+1$, monotonicity keeps an alternative boundary between the next two reference boundaries. A boundary reaching either endpoint has joined the reference parsing.}
\label{fig:boundaries}
\end{figure}
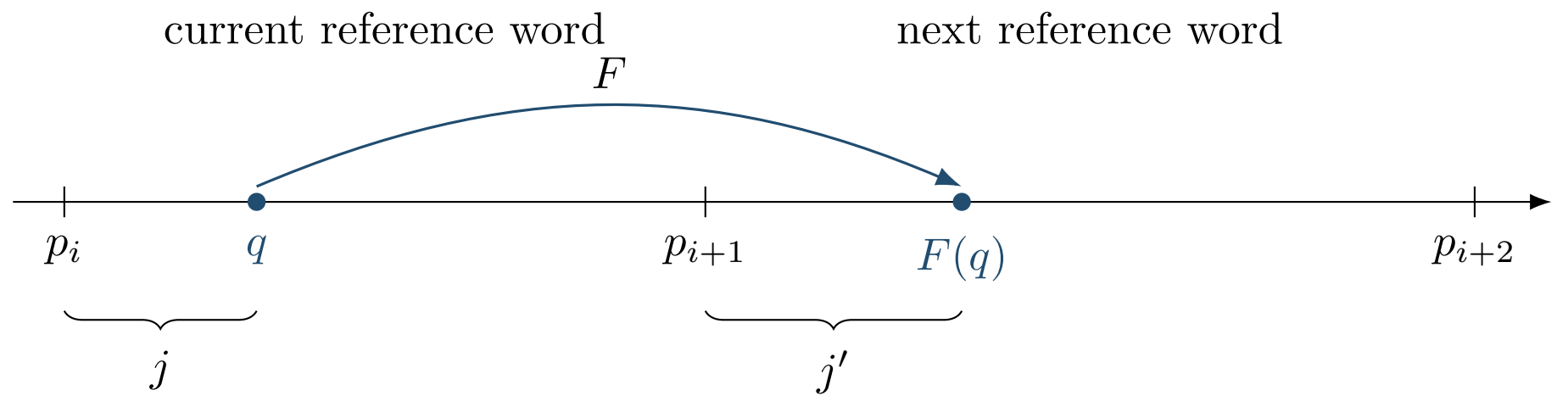

\subsection{When every splice remains short}

Use zero-based slices: $u[x:r]$ is the suffix of a word $u\in\Sig^r$ after position $x$, and $u[0:x]$ is its prefix of length $x$. For $1\le x<r$, call $A$ \emph{closed at $x$} if
\begin{equation}\label{eq:closed}
 u[x:r]v[0:x]\in A\qquad\text{for all }u,v\in A.
\end{equation}
This is precisely the situation in which every length-$r$ window at offset $x$ across two short reference words is itself short.

\begin{lemma}\label{lem:product}
If $A$ is closed at $x$ and $g=\gcd(x,r)$, then
\[
 A=D^{r/g}
\]
for a nonempty proper subset $D\subsetneq\Sig^g$. Conversely, such a product is closed at every offset divisible by $g$.
\end{lemma}
\begin{proof}
Taking $u=v$ in~\eqref{eq:closed} shows that rotation by $x$ maps $A$ into itself. Since $A$ is finite, it is invariant under this rotation and its inverse. The rotations generated by $x$ modulo $r$ are exactly the multiples of $g$, so rotation by $g$ also preserves $A$.

Rotating the splice in~\eqref{eq:closed} back by $x$ gives
\[
 v[0:x]u[x:r]\in A.
\]
Thus the first $x$ coordinates of one word can be copied independently into another word of $A$. Say that a set of coordinates $S$ is \emph{replaceable} if every word formed from one member of $A$ on $S$ and another on its complement still belongs to $A$. Replaceable sets are closed under complement by interchanging the two words. They are closed under intersection as well: first copy $u$ into $S$ of $v$, then copy the resulting word into $T$ of a fresh copy of $v$. The final word uses $u$ exactly on $S\cap T$.

Partition the coordinates into $m=r/g$ consecutive blocks of length $g$, and put $\ell=x/g$. The first $\ell$ blocks are replaceable. Rotation invariance makes every cyclic interval of $\ell$ blocks replaceable. These intervals separate distinct blocks: the incidence patterns of two distinct positions in all proper, nonempty cyclic intervals of a fixed length are different. One way to see this is that the indicator of a single proper cyclic interval has exactly one change from $0$ to $1$; it cannot be invariant under a nonzero rotation. Therefore the Boolean algebra generated by the intervals contains each individual block. Each block is replaceable.

Let $D_i$ be the projection of $A$ onto block $i$. Replacing one block at a time shows that all choices from the projections are independent, so $A=D_1\times\cdots\times D_m$. Rotation by $g$ permutes the blocks and preserves $A$, forcing all $D_i$ to be the same set $D$. Since $A$ is nonempty and proper, so is $D$. The converse follows because a splice at a multiple of $g$ joins whole blocks belonging to $D$.
\end{proof}

\subsection{Advancing the least remaining offset}

At the end of a round, the last reference word is a short word $c\in A$. Any parsing that has not met the reference has its most recent boundary at an interior offset of $c$. We append two codewords, the last of which is again short, and follow these boundaries through two applications of $F$.

\begin{lemma}\label{lem:round}
Suppose the least surviving offset in $c\in A$ is $x$, where $1\le x<r$. There are two codewords, the second short, whose concatenation has length at most $2r+1$ and leaves every surviving offset at least $x+1$. If $A$ is not closed at $x$, the length is at most $2r$.
\end{lemma}
\begin{proof}
Suppose first that closure fails. Choose $u,v\in A$ with
$u[x:r]v[0:x]\notin A$, and append $uv$. In moving from $c$ to $u$, an offset $x$ either increases or stays fixed, by~\eqref{eq:offset}. If it increases, the next transition cannot decrease it, since $u$ is short. If it stays fixed, the window at offset $x$ across $u,v$ lies outside $A$, so the alternative word has length $r+1$ and its offset increases at the next transition. An endpoint is removed as soon as it meets the reference. Monotonicity of $F^2$ gives the same lower bound for all larger surviving offsets. The cost is $2r$.

Now suppose closure holds. By Lemma~\ref{lem:product}, write
\[
 g=\gcd(x,r),\quad m=r/g,\quad\ell=x/g,\quad A=D^m.
\]
Choose $p\in D$, $q\in\Sig^g\setminus D$, and $y\in\Sig$. Form
\begin{equation}\label{eq:longword}
 L=q\,p^{\ell-1}\,y\,q\,p^{m-\ell-1}.
\end{equation}
Its length is $r+1$, and its first length-$g$ block is $q\notin D$, so its length-$r$ prefix is outside $A$. Thus $L\in C(A)$. Append $L$ followed by any $d\in A$.

Place $c$, $L$, and $d$ on the half-open intervals $[-r,0)$, $[0,r+1)$, and $[r+1,2r+1)$, respectively. The boundary at $-r+x$ sees the length-$r$ window
\[
 c[x:r]L[0:x]=c[x:r]q\,p^{\ell-1}.
\]
It contains the forbidden block $q$ at an aligned block position, and hence lies outside $A$. The alternative word is long, ending at $x+1$ inside $L$. Its next window is
\[
 L[x+1:r+1]d[0:x]=q\,p^{m-\ell-1}d[0:x],
\]
which again contains an aligned forbidden block. The next alternative word is long too, and ends at $r+x+2$, at offset $x+1$ in $d$. If $x+1=r$, it meets the right endpoint of $d$. Monotonicity of $F^2$ controls the larger offsets. The appended length is $2r+1$.
\end{proof}

\begin{figure}[htb]
\centering
\begin{tikzpicture}[x=.87cm,y=1cm]
 \draw[fill=black!3] (0,0) rectangle (3,.65);
 \draw[fill=lightblue] (3,0) rectangle (8,.65);
 \draw[fill=black!3] (8,0) rectangle (11,.65);
 \node at (1.5,.33) {$c\in D^m$};
 \node at (5.5,.33) {$L=q\,p^{\ell-1}\,y\,q\,p^{m-\ell-1}$};
 \node at (9.5,.33) {$d\in D^m$};
 \node[below] at (0,0) {$-r$};\node[below] at (3,0) {$0$};
 \node[below] at (8,0) {$r+1$};\node[below] at (11,0) {$2r+1$};
 \fill[inkblue] (1.3,.85) circle(2pt);
 \fill[inkblue] (5,.85) circle(2pt);
 \fill[inkblue] (9.4,.85) circle(2pt);
 \draw[flow] (1.3,.85) to[bend left=20] node[above]{$r+1$}(5,.85);
 \draw[flow] (5,.85) to[bend left=20] node[above]{$r+1$}(9.4,.85);
 \node[above] at (1.3,1.45) {$-r+x$};
 \node[above] at (5,1.45) {$x+1$};
 \node[above] at (9.4,1.45) {$r+x+2$};
\end{tikzpicture}
\caption{The long--short round. The two copies of $q\notin D$ force two long alternative codewords. Their boundaries advance from offset $x$ in $c$ to offset $x+1$ in $d$. Positions are schematic; the displayed coordinates give the exact lengths.}
\label{fig:round}
\end{figure}
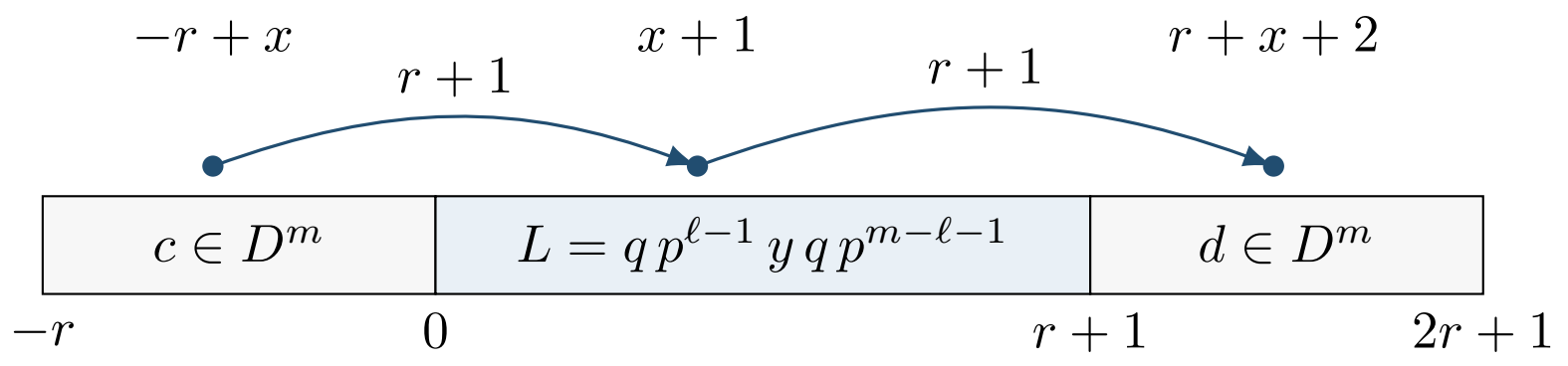

\begin{proof}[Proof of the upper bound in Theorem~\ref{thm:two}]
If $r=1$, any letter in $A$ resets the decoder, and the threshold is one because $A$ is proper and the decoder has a nonroot state.

Let $r\ge2$. Begin with any $c\in A$. Every initial state completes a codeword during these $r$ letters, so Lemma~\ref{lem:suffix} shows that every output is a suffix of $c$ of length at most $r-1$. The root's parsing ends at the root and serves as the reference. Every other output corresponds to a boundary at an interior offset in $c$.

Apply Lemma~\ref{lem:round} to the least surviving offset. Its conclusion raises that offset by at least one and leaves the last reference word short. Repeat until no offset remains. This takes at most $r-1$ rounds, since the possible offsets are $1,\ldots,r-1$. At each round end, an alternative parsing either has met the reference or has its last boundary in the last short word: its next word has length at least $r$, so it cannot complete again strictly inside that word. Thus the same representation applies at every round. When all alternatives have met the reference, the input resets to $\eps$.

The total length is at most
\[
 r+(r-1)(2r+1)=2r^2-1.
\]
All windows used in the construction lie in the chosen words. An arbitrary infinite continuation used to define $F$ cannot affect them. Polynomial-time implementation is discussed in Section~\ref{sec:algorithms}.
\end{proof}

\subsection{Equality and changes of alphabet}

\begin{lemma}\label{lem:lift}
Let $\theta:\Sig\twoheadrightarrow\Gamma$ be a surjective letter map, let $C'\subseteq\Gamma^+$ be a finite maximal prefix code, and set $C=\theta^{-1}(C')$. Then $C$ is a finite maximal prefix code. A word $w$ resets its literal decoder if and only if $\theta(w)$ resets the decoder of $C'$. Consequently the thresholds agree whenever the decoders synchronize.
\end{lemma}
\begin{proof}
The state set is $Q=\theta^{-1}(Q')$, and the induced map $Q\to Q'$ is onto and respects transitions. Prefix-freeness and fullness are preserved, so $C$ is a finite maximal prefix code. A reset word projects to a reset word.

Conversely, if $\theta(w)$ resets, all states in $Q\cdot w$ have the same image, and thus the same length $h$. If one initial state completes a codeword while reading $w$, Lemma~\ref{lem:suffix} gives $h<|w|$. Any thread that completes none would end at length $|q|+|w|\ge|w|$, a contradiction. Hence all threads complete, and all outputs are the same length-$h$ suffix of $w$.

If no thread completes, all output lengths are $|q|+|w|$. They cannot be equal when $Q$ contains both the root and a nonroot state. The remaining case $Q=\{\eps\}$ is immediate. This proves the converse. Every word over $\Gamma$ has a lift of the same length, giving equality of thresholds.
\end{proof}

\begin{proof}[Proof of the equality assertion in Theorem~\ref{thm:two}]
Suppose $r\ge2$ and $A$ is not a subalphabet power. Then $A$ is not closed at $r-1$: otherwise Lemma~\ref{lem:product}, with $\gcd(r-1,r)=1$, would give $A=T^r$.

If the construction uses $r-1$ rounds, their least offsets must be $1,2,\ldots,r-1$. The last round costs at most $2r$, saving one letter from the bound. If it uses fewer rounds, the saving is larger. Thus
\[
 s(C(A))\le2r^2-2.
\]
Conversely, if $A=T^r$, send the letters of $T$ to $a$ and all remaining letters to $b$. Then $C(A)$ is the inverse image of $C_{r+1}$. Lemma~\ref{lem:lift} and Theorem~\ref{thm:lower}, proved in the next section, give $s(C(A))=2r^2-1$. For $r=1$, every nonempty proper $A$ is itself a subalphabet, and the threshold is one.
\end{proof}

\section{An exact extremal family}\label{sec:extremal}

Put $n=k-1\ge1$ and consider
\[
 C_k=\{a^n\}\cup\bigl(\{a,b\}^{n+1}\setminus a^n\{a,b\}\bigr).
\]
The states of its decoder are
\[
 Q=\{q\in\{a,b\}^*:|q|\le n\}\setminus\{a^n\}.
\]
Every word of length less than $n$ is an internal vertex; at depth $n$, only $a^n$ is a leaf. Thus
\begin{equation}\label{eq:binarytransition}
 q\cdot x=\begin{cases}
 \eps,&|q|=n\text{ or }(q=a^{n-1}\text{ and }x=a),\\
 qx,&\text{otherwise}.
 \end{cases}
\end{equation}

\subsection{A reset word of length \texorpdfstring{$2n^2-1$}{2n\textasciicircum 2-1}}

\begin{lemma}\label{lem:binaryupper}
The word $W=a^n(b^{n+1}a^n)^{n-1}$ resets $Q$ to $\eps$.
\end{lemma}
\begin{proof}
Let $Z=\{\eps,a,\ldots,a^{n-1}\}$. Reading $a^n$ fixes each state of $Z$. From a state containing $b$, of length $j$, the first $n+1-j$ letters complete a long codeword and the remaining $j-1$ letters leave $a^{j-1}$. Hence $Q\cdot a^n=Z$.

For $0\le i<n$, reading $b^{n+1}$ from $a^i$ completes a long word after $n+1-i$ letters and leaves $b^i$. Reading $a^n$ from $b^i$ then leaves $a^{i-1}$ if $i\ge1$, and leaves $\eps$ if $i=0$. The block $V=b^{n+1}a^n$ therefore acts on $Z$ by
\[
 a^i\cdot V=a^{\max(i-1,0)}.
\]
After $n-1$ copies, every state reaches $\eps$. The length is
$n+(n-1)(2n+1)=2n^2-1$.
\end{proof}

For the converse, a reset word may end at any state. We must bound all such words, not just those returning to $\eps$. Once the states are suffixes of a common input, their lengths determine them. We use these lengths to follow every possible reset word.

\subsection{Ages and the first short completion}

Call the length of a suffix state its \emph{age}. Ages lie in $\{0,\ldots,n\}$. On this set define
\[
 \sigma(g)=g+1\pmod{n+1},\qquad
 \varphi(g)=\begin{cases}
 g+1,&0\le g\le n-2,\\
 0,&g=n-1\text{ or }g=n.
 \end{cases}
\]
On the smaller circle $\{0,\ldots,n-1\}$, let $\rho(g)=g+1\pmod n$.

Suppose the current states are suffixes of a common word $P$ of length at least $n$. Let $c(P)$ be the trailing run of $a$'s, capped at $n$. A letter $b$ acts on ages by $\sigma$ and sets the counter to zero. A letter $a$ acts by $\sigma$ if the prior counter is at most $n-2$, and by $\varphi$ if it is at least $n-1$. In the latter case the counter becomes $n$. Call such an $a$ a \emph{special letter}.

These rules are exact. Age $n$ always returns to zero on the next letter. Age $n-1$ returns to zero on $a$ exactly when its suffix is $a^{n-1}$, which is equivalent to the common input having at least $n-1$ trailing $a$'s. Every other transition increases age by one. In each case the new state is the suffix of the enlarged input of the new age. Distinct ages represent distinct states.

Two initial state sets fit this description:
\[
 Z=\{\eps,a,\ldots,a^{n-1}\},\qquad
 Z_b=\{\eps,b,\ldots,b^n\}.
\]
For $Z$, take the common context $a^n$, all small-circle ages, and counter $n$. For $Z_b$, take the context $b^{n+1}$, all big-circle ages, and counter zero. These contexts represent the sets; they are not charged as part of a subsequent input.

Let $d(Z)$ be the minimum length needed to send $Z$ to a singleton. It is finite by Lemma~\ref{lem:binaryupper}.

\begin{lemma}\label{lem:mandatory}
The reset threshold satisfies $s(C_k)=n+d(Z)$.
\end{lemma}
\begin{proof}
Since $Q\cdot a^n=Z$, we have $s(C_k)\le n+d(Z)$. Conversely, let $w$ reset $Q$. It also resets $Z_b\subseteq Q$. Until the first special letter, the age action on $Z_b$ is a permutation of all $n+1$ ages. A special letter must therefore occur.

Let $t$ be the number of letters consumed up to and including that first special letter. Because the initial counter is zero, the consumed prefix ends with $a^n$, so $t\ge n$. Immediately after this letter, the age set is exactly $\{0,\ldots,n-1\}$ and the counter is $n$. The actual states are the corresponding suffixes of $a^n$, namely $Z$. The remaining suffix of $w$ must synchronize $Z$, and has length at least $d(Z)$. Thus $|w|\ge t+d(Z)\ge n+d(Z)$.
\end{proof}

For $n=1$, $Z=\{\eps\}$ and $d(Z)=0$, so the threshold is one. Assume from now on that $n\ge2$.

\subsection{Excursions away from the small circle}

After a special letter the ages lie on the small circle and the counter equals $n$. Further special letters are $a$'s, acting as $\rho$. To leave such a run requires a $b$. Before the next special letter there must be $n-1$ ordinary $a$'s after the last $b$. Thus the intervening block $B$ of nonspecial letters starts with $b$, ends with $ba^{n-1}$, and has length $j\ge n$.

Suppose this block is followed by $e\ge1$ special letters. Put $\tau=j\bmod(n+1)$. Let $\iota$ include the small circle in the big one, and let $\pi$ fix $0,\ldots,n-1$ and send $n$ to $n-1$. Since $\varphi=\rho\circ\pi$, the whole excursion acts on small-circle ages as
\[
 \rho^e\circ\pi\circ\sigma^j\circ\iota.
\]
For $0\le m\le n-2$, define
\begin{equation}\label{eq:slip}
 X_m(g)=\begin{cases}
 g-m\pmod n,&0\le g\le m,\\
 g-m-1\pmod n,&m+1\le g\le n-1.
 \end{cases}
\end{equation}
The only nontrivial fiber of $X_m$ is $\{m,m+1\}$, which maps to zero. Its image omits $h_m=n-m-1$.

\begin{figure}[htb]
\centering
\begin{tikzpicture}[x=1.35cm,y=1cm]
 \foreach \i in {0,...,4} {\node[vertex] (s\i) at (\i,1.5) {$\i$};}
 \node[marked] at (1,1.5) {$1$};\node[marked] at (2,1.5) {$2$};
 \node[vertex] (t4) at (0,0) {$4$};
 \node[marked] (t0) at (1.5,0) {$0$};
 \node[vertex] (t1) at (3,0) {$1$};
 \node[vertex] (t2) at (4,0) {$2$};
 \node[vertex,dashed,draw=black!45,text=black!55] (t3) at (5.2,0) {$3$};
 \draw[->] (s0)--(t4);
 \draw[flow] (s1)--(t0);\draw[flow] (s2)--(t0);
 \draw[->] (s3)--(t1);\draw[->] (s4)--(t2);
 \node[left] at (-.65,1.5) {before};\node[left] at (-.65,0) {after};
 \node[below] at (5.2,-.4) {omitted age};
\end{tikzpicture}
\caption{The slip $X_1$ on five ages. The ages $1$ and $2$ merge at $0$, and age $3$ is omitted. The target circle is cut immediately after $3$ and displayed in the order $4,0,1,2,3$. This makes the increasing lift used in Appendix~\ref{app:arcs} visible.}
\label{fig:slip}
\end{figure}
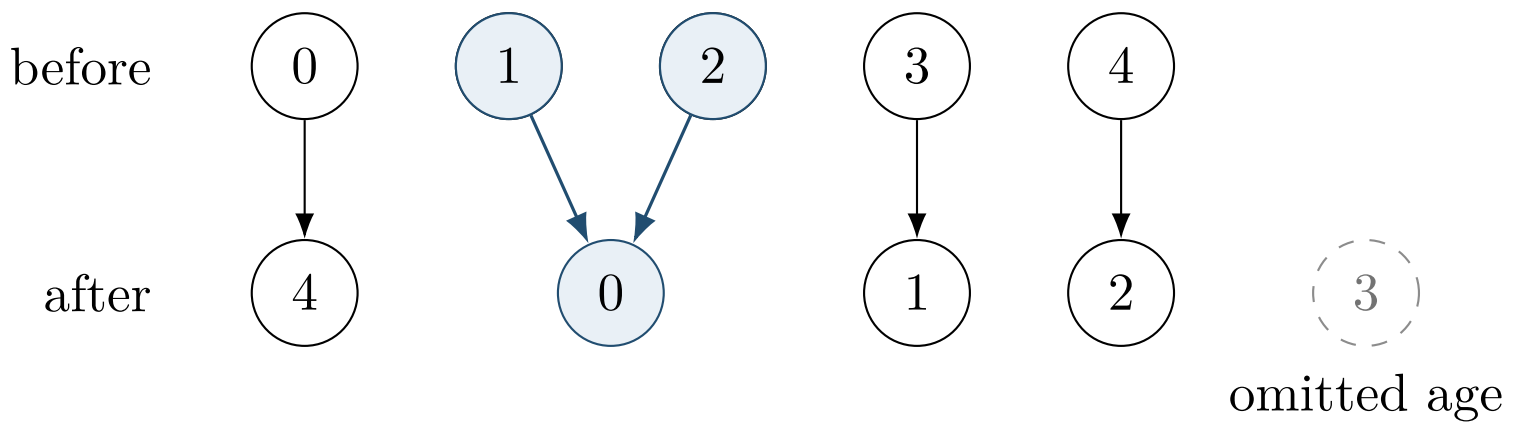

Splitting $g$ according to whether $g+\tau$ crosses $n$ or $n+1$ gives the following three possibilities:
\begin{center}
\begin{tabular}{@{}lll@{}}
\toprule
$\tau$ & Excursion map & Necessary length\\
\midrule
$0$ & $\rho^e$ & $j+e\ge n+1+e$\\
$n$ & $\rho^{e-1}$ & $j+e\ge n+e$\\
$1\le\tau\le n-1$ & $\rho^{e-1}\circ X_{n-1-\tau}$ & $j+e\ge n+1+\tau+e$\\
\bottomrule
\end{tabular}
\end{center}
For completeness, in the middle range put $m=n-1-\tau$. If $g<m$, the big-circle age $g+\tau$ is at most $n-2$ and the first special letter sends it to $g+\tau+1=g-m+n$. If $g=m$ or $m+1$, it sends it to zero. If $g>m+1$, reduction modulo $n+1$ occurs before applying $\varphi$, and the result is $g-m-1$. These are exactly~\eqref{eq:slip}. The remaining $e-1$ letters rotate. The cases $\tau=0,n$ follow directly from $\varphi\iota=\rho$ and $\varphi\sigma^n\iota=\mathrm{id}$.

Assign cost $1$ to $\rho$ and cost $2n+1-m$ to $X_m$. Each excursion can be replaced by these operations at no greater cost. The first two rows require only $e$ or $e-1$ rotations. In the third row,
\begin{equation}\label{eq:correctcost}
 (2n+1-m)+(e-1)=n+1+\tau+e\le j+e,
 \qquad m=n-1-\tau.
\end{equation}
Here $j\ge n$ and $1\le\tau<n$ imply $j\ge n+1+\tau$.

An initial run of special letters is already a sequence of rotations. A final unfinished block of nonspecial letters acts by a power of $\sigma$, so it cannot change cardinality and can be discarded when testing synchronization. Consequently, if $D_{\mathrm{slip}}(G)$ is the minimum total cost of rotations and slips sending $G$ to a singleton, then
\begin{equation}\label{eq:comparison}
 d(Z)\ge D_{\mathrm{slip}}(\Z/n\Z).
\end{equation}
We only need this direction: allowing additional sequences of slips can lower the minimum, which is harmless for~\eqref{eq:comparison}.

\subsection{A potential on circular arcs}

Cardinality alone does not give the sharp cost: slips at different positions have different prices. We also keep track of where a shortest covering arc starts.

For $1\le p\le n$ and $1-p\le\alpha\le n-p$, let
\[
 I(\alpha,p)=\{\alpha,\alpha+1,\ldots,\alpha+p-1\}\pmod n,
 \qquad \Psi(\alpha,p)=(p-1)(2n+1)-\alpha.
\]
For $p<n$, this range selects a unique integer representative of the starting point of each arc. For $p=n$, all starts describe the full circle, and the least value occurs at $\alpha=0$.

For a nonempty subset $G$ of the circle, put
\begin{equation}\label{eq:potential}
 \Phi(G)=\begin{cases}
 0,&|G|=1,\\
 \min\{\Psi(\alpha,p):2\le p\le n,\ G\subseteq I(\alpha,p),\
           1-p\le\alpha\le n-p\},&|G|\ge2.
 \end{cases}
\end{equation}
In particular,
\begin{equation}\label{eq:fullpotential}
 \Phi(\Z/n\Z)=(n-1)(2n+1).
\end{equation}

\begin{lemma}\label{lem:hull}
An arc of smaller size has smaller $\Psi$ than every arc of larger size. For $2\le|G|<n$, the minimum in~\eqref{eq:potential} selects a unique covering arc of minimum size. Its size is at most $n-1$, and both its endpoints lie in $G$.
\end{lemma}
\begin{proof}
At size $p$, the maximum possible value is $(p-1)(2n+2)$. The minimum at size $p+1$ is $p(2n+1)-(n-p-1)$, larger by $n+3$. Iterating gives the first assertion. A proper subset is covered by an arc of size $n-1$, obtained by omitting a point outside it. Among the minimum-size covers, different canonical starts give different potentials, so one is selected uniquely. If an endpoint were absent from $G$, deleting it would give a smaller cover. The only possible reduction to a singleton would contradict $|G|\ge2$.
\end{proof}

Write $\hull(G)$ for this selected arc. The word ``selected'' matters: a set may have several shortest covering arcs, and $\Psi$ resolves that choice.

\begin{lemma}\label{lem:potential}
For $|G|\ge2$ and $0\le m\le n-2$,
\begin{align}
 \Phi(G)&\le1+\Phi(\rho G),\label{eq:rotationpotential}\\
 \Phi(G)&\le2n+1-m+\Phi(X_mG).\label{eq:slippotential}
\end{align}
\end{lemma}
\begin{proof}
For rotation, the full circle is immediate. Otherwise reverse-rotate $\hull(\rho G)=I(\alpha,p)$. If $\alpha>1-p$, its canonical start decreases by one and its potential increases by one. At $\alpha=1-p$, the new canonical start is $n-p$, and the potential changes by $1-n$. The reverse-rotated arc covers $G$, proving~\eqref{eq:rotationpotential}.

If $X_mG$ is a singleton, the only possible $G$ is $\{m,m+1\}$. This is the arc $I(m,2)$ and has potential $2n+1-m$, including the case $n=2,m=0$ of the full circle. Thus~\eqref{eq:slippotential} holds with equality.

Suppose $|X_mG|\ge2$. Its image omits $h=n-m-1$, so its selected covering arc $I_0=\hull(X_mG)$ is proper. Neither endpoint is $h$, by Lemma~\ref{lem:hull}. The preimage $P=X_m^{-1}(I_0)$ is a circular arc of size at least two, and
\begin{equation}\label{eq:arcpreimage}
 \Psi(P)-\Psi(I_0)\le2n+1-m.
\end{equation}
The calculation is given in Appendix~\ref{app:arcs}. It uses the increasing lift of $X_m$ obtained by cutting the target circle at its omitted point. An interval avoiding that point lifts to an interval; one crossing it has preimage formed by the two end intervals, which together give a wrapping arc. Since $P$ covers $G$, inequality~\eqref{eq:arcpreimage} proves~\eqref{eq:slippotential}.
\end{proof}

\begin{proof}[Proof of Theorem~\ref{thm:lower}]
The case $n=1$ follows from Lemma~\ref{lem:mandatory}. For $n\ge2$, take any sequence of rotations and slips from the full small circle to its first singleton. At each step Lemma~\ref{lem:potential} bounds the decrease of $\Phi$ by the cost of that step. Summing gives total cost at least
\[
 \Phi(\Z/n\Z)-0=(n-1)(2n+1).
\]
By~\eqref{eq:comparison} and Lemma~\ref{lem:mandatory},
\[
 s(C_k)=n+d(Z)\ge n+(n-1)(2n+1)=2n^2-1.
\]
Lemma~\ref{lem:binaryupper} gives equality. For any alphabet with at least two letters, a surjective map onto $\{a,b\}$ and Lemma~\ref{lem:lift} transfer the exact threshold.
\end{proof}

\begin{proof}[Proof of Corollary~\ref{cor:growth}]
Theorem~\ref{thm:universal} gives the upper bound. The lifted binary family gives the lower bound at every height over each fixed alphabet. Both are quadratic with positive leading coefficient.
\end{proof}

\section{Constructing the reset words}\label{sec:algorithms}

Both proofs give algorithms on an explicit prefix tree. The short-detection proof uses infinite series only to establish the length bound; the algorithm itself uses finite rational linear algebra.

\subsection{A basis of short observations}

Let $N=|Q|$, $d=|\Sig|$, and $h=2k-3$. For each letter $b$, let $B_b$ be the deterministic transition matrix, with $(B_b)_{pq}=1$ when $p\cdot b=q$. Set $B_w=B_{b_1}\cdots B_{b_t}$ for $w=b_1\cdots b_t$, and let $\alpha$ be the row vector that is one on $Z$ and zero elsewhere. For $\varnothing\ne S\subsetneq Z$, define the column vector
\[
 \nu_S(q)=r\one_S(\pi(q))-|S|.
\]
Then
\begin{equation}\label{eq:matrixobs}
 \alpha B_w\nu_S=r\bigl(|T_w^{-1}(S)|-|S|\bigr).
\end{equation}
We can test all words of length at most $h$ by testing a basis of their row vectors.

\begin{lemma}\label{lem:bfs}
There is an algorithm that constructs at most $N$ words $\mathcal B\subseteq\Sig^{\le h}$ such that the rows $\alpha B_w$, $w\in\mathcal B$, form a basis of
\[
 V_h=\Span_{\Q}\{\alpha B_w:|w|\le h\}.
\]
It tests at most $1+dN$ candidate rows for independence.
\end{lemma}
\begin{proof}
Start with $\alpha$, represented by $\eps$. Process retained rows in nondecreasing word length. For each retained word $w$ with $|w|<h$, test the rows represented by $wb$, $b\in\Sig$, retaining just those independent of the rows already retained. At most $N$ rows can be retained, and each has at most $d$ tested extensions.

To check that discarding a row loses nothing, induct on length. Suppose every row represented by a word of length at most $t$ lies in the span of retained rows whose represented lengths are at most $t$. Multiplying such an expression by $B_b$ expresses each length-$(t+1)$ row in terms of extensions of those retained rows. These extensions are tested by depth $t+1$ and either retained or expressed in previously retained rows of at most that depth. This proves the assertion up to the cutoff $h$.
\end{proof}

\begin{theorem}\label{thm:algorithm}
Given an explicitly listed finite maximal prefix code of height $k\ge2$ and its alphabet, one can decide synchronization and, when it holds, construct a reset word of length at most $(k-1)(4k-5)$ in polynomial bit complexity.
\end{theorem}
\begin{proof}
Construct the tree and transition table, choose a letter $a$, and find its pure leaf $a^r$. If $r=1$, output $a^{k-1}$. Otherwise compute the basis $\mathcal B$ of Lemma~\ref{lem:bfs} once.

For each set $S$ encountered in the preimage construction, evaluate $\alpha B_w\nu_S$ for $w\in\mathcal B$. If all values vanish, then~\eqref{eq:matrixobs} vanishes for every word of length at most $h$. Theorem~\ref{thm:short} makes it vanish for all words. A reset word would give a nonzero value, as in Lemma~\ref{lem:unbalance}, so this certifies nonsynchronization.

Otherwise a basis word supplies the required short unbalancing word. For the first singleton test, choose a fiber of size at least two. At later steps, test the $r$ rotations to find a larger preimage. There are at most $r-1$ expansions. If the procedure does not return a nonsynchronization certificate, it outputs~\eqref{eq:reset}. Proposition~\ref{prop:cycle} proves that this word resets and satisfies the length bound. Thus both answers are sound, and the procedure terminates on every input.

For the bit bound, entries of $\alpha B_w$ count the images of $r$ starting states, so they are integers between $0$ and $r$. There are at most $1+dN$ candidate rows of dimension $N$. Rational Gaussian elimination can be implemented with intermediate coordinates given by ratios of minors of the integer input matrix. A minor of order $s\le N$ has absolute value at most $(r\sqrt{s})^s$ by Hadamard's inequality. Numerators and denominators therefore have $O(N\log(Nr+1))$ bits; the same polynomial estimate applies to the actual residual and normalized pivot coordinates. Integer arithmetic and reduction of fractions then take polynomial time. The number of arithmetic operations, row evaluations, rotations, and transition simulations is polynomial in $N,d,k$. The output has $O(k^2)$ letters. All these parameters are bounded by the explicit alphabet and code listing size.
\end{proof}

At height one the decoder is a singleton and the algorithm returns $\eps$. The procedure constructs a reset word within the bound; it does not compute a shortest reset word in general.

\subsection{The consecutive-depth construction}

Lemma~\ref{lem:round} can be implemented directly. Test closure at $x$ by examining the pairs $u,v\in A$. A failed test supplies the short--short round. If closure holds, compute $g=\gcd(x,r)$ and project $A$ onto a block of length $g$ to obtain $D$. Choose $p\in D$ and enumerate $\Sig^g$ to find $q\notin D$. The long--short round is then given by~\eqref{eq:longword}.

These searches are polynomial in the input size. Indeed,
\[
 |\Sig|^g\le |\Sig|^r\le |C(A)|,
\]
since every word of length $r$ contributes either one short leaf or $d$ long leaves. There are at most $r-1$ rounds, and all decoder images and offsets can be updated by direct simulation.

An equivalent implementation avoids computing $D$: enumerate pairs $(u,v)\in C(A)\times A$ and test whether the pair advances all currently eligible offsets. The forcing lemma guarantees a successful pair of total length at most $2r+1$. When closure fails, short--short pairs already suffice. This is the finite-search formulation used by the verified implementation.

\section{The remaining gap}\label{sec:discussion}

Does every synchronizing finite maximal prefix code of height $k\ge2$ satisfy
\[
 s(C)\le2(k-1)^2-1?
\]
The consecutive-depth theorem and the binary family show that this would be the exact bound over every alphabet with at least two letters. Corollary~\ref{cor:shortcycle} also proves it when a sufficiently short pure-letter codeword is present, under the synchronization hypothesis.

The two upper-bound proofs indicate where the gap arises. In the general construction, each preimage enlargement requires a short unbalancing word, a return to the pure-letter cycle, and possibly a rotation. In the consecutive-depth construction, order along the input lets one control all alternative parsings by the least surviving offset. This avoids the repeated cycle-compression cost.

That order need not hold in an arbitrary prefix tree. If the next codeword length at position $j$ is $\ell_j$, then
\[
 F(j+1)-F(j)=1+\ell_{j+1}-\ell_j,
\]
which can be negative. Thus the proof for two lengths cannot simply be applied after replacing the other lengths by consecutive ones. A sharper general bound would need either a cheaper preimage construction or a way to control parsings without this monotonicity.

\begin{example}
For $k=3$, the extremal code is
\[
 C_3=\{aa,aba,abb,baa,bab,bba,bbb\},
\]
and $aa\,bbb\,aa$ is a shortest reset word, of length seven. In contrast, $A=\{aa,ab\}$ gives
\[
 C(A)=\{aa,ab,baa,bab,bba,bbb\}.
\]
Here $A$ is not a subalphabet square, so Theorem~\ref{thm:two} gives the strict upper bound $s(C(A))\le6$.
\end{example}

\appendix
\section{Preimages of arcs under a slip}\label{app:arcs}

We prove~\eqref{eq:arcpreimage}, completing Lemma~\ref{lem:potential}. Fix $0\le m\le n-2$ and let $h=n-m-1$. Let $I_0=I(\alpha_0,p_0)$ be a proper arc with $2\le p_0\le n-1$, neither endpoint equal to $h$. The increasing lift
\[
 \widehat X_m:\{0,\ldots,n-1\}\longrightarrow
 L=\{-m,-m+1,\ldots,n-m-2\}
\]
is defined by
\[
 \widehat X_m(g)=\begin{cases}g-m,&g\le m,\\g-m-1,&g\ge m+1.\end{cases}
\]
It takes every value in $L$ once except zero, which has preimage $\{m,m+1\}$. Reduction modulo $n$ gives $X_m$. The single missing residue is $h$.

\paragraph{The arc avoids $h$.}
There is a unique lift $[u,u+p_0-1]\subseteq L$ of $I_0$. Its preimage is an ordinary integer interval. The three possible positions of this interval relative to zero give Table~\ref{tab:avoid}. All displayed starts of $P$ are canonical. In the first row the start of $I_0$ is $u$; in the second it is also $u$; in the third it is $u+n$. Substitution into $\Psi(\alpha,p)=(p-1)(2n+1)-\alpha$ gives the stated differences. If the first row gives $p_0+1=n$, the preimage is the full circle and its displayed start is necessarily zero.

\begin{table}[htb]
\centering
\renewcommand{\arraystretch}{1.25}
\begin{tabular}{@{}lll@{}}
\toprule
Position of the lifted interval & $P=X_m^{-1}(I_0)$ & $\Psi(P)-\Psi(I_0)$\\
\midrule
$u\le0\le u+p_0-1$ & $I(u+m,p_0+1)$ & $2n+1-m$\\
$u\ge1$ & $I(u+m+1,p_0)$ & $-m-1$\\
$u+p_0-1\le-1$ & $I(u+m,p_0)$ & $n-m$\\
\bottomrule
\end{tabular}
\caption{The preimage of an arc avoiding the omitted point. The first case gains one point by containing the doubled fiber.}
\label{tab:avoid}
\end{table}

\paragraph{The arc contains $h$.}
Since $h$ is not an endpoint, write
\[
 I_0=\{h-t,\ldots,h,\ldots,h+s\}\pmod n,
 \qquad t,s\ge1,\qquad p_0=t+s+1\le n-1.
\]
The part after $h$ lifts to $[-m,-m+s-1]$, and the part before $h$ lifts to $[n-m-1-t,n-m-2]$. Their preimages are an initial and a final integer interval in $\{0,\ldots,n-1\}$. The gap has a nonempty preimage because $p_0\le n-1$, so their union is a proper wrapping arc.

There are three cases, according to which lifted interval contains zero; see Table~\ref{tab:cross}. The first two cannot occur together: $s\ge m+1$ and $t\ge h$ would imply $t+s+1\ge n+1$. The implications in the first column follow from $t+s+1\le n-1$. The canonical start of $I_0$ is $h-t-n$ in the first row and $h-t$ in the other two. The displayed starts of $P$ are canonical as well. These values give the potential differences in the table.

\begin{table}[htb]
\centering
\renewcommand{\arraystretch}{1.25}
\begin{tabular}{@{}lll@{}}
\toprule
Position relative to the doubled fiber & $P=X_m^{-1}(I_0)$ & $\Psi(P)-\Psi(I_0)$\\
\midrule
$s\ge m+1$ (so $t<h$) & $I(-t,p_0)$ & $-m-1$\\
$t\ge h$ (so $s\le m-1$) & $I(-1-t,p_0)$ & $n-m$\\
$s\le m$, $t\le h-1$ & $I(-t,p_0-1)$ & $-n-m-2$\\
\bottomrule
\end{tabular}
\caption{The preimage of an arc crossing the omitted point. In the last case the arc loses the point $h$ and contains no doubled fiber, so its size decreases by one.}
\label{tab:cross}
\end{table}

The preimage has size at least two in every case. In the last row this follows from $p_0=t+s+1\ge3$. All six differences are at most $2n+1-m$. This proves~\eqref{eq:arcpreimage}.

\FloatBarrier
\section{Formal verification and reproducibility}\label{app:lean}

The Lean development uses proper-prefix states and proves all four main results, including short detection, the parsing and forcing lemmas, the age reduction, and the arc-potential inequality.

The main results are collected in \texttt{HeightSync.Manuscript}:
\begin{center}
\begin{tabular}{@{}ll@{}}
\toprule
Paper statement & Lean declaration\\
\midrule
Theorem~\ref{thm:universal} & \texttt{universal}\\
Theorem~\ref{thm:two} & \texttt{consecutive}\\
Theorem~\ref{thm:lower} & \texttt{binary}\\
Corollary~\ref{cor:growth} & \texttt{growth}\\
\bottomrule
\end{tabular}
\end{center}
The complete statement index is provided in \texttt{MANUSCRIPT\_COVERAGE.md}. Short detection is formalized over arbitrary fields. For the embedding lemma, the development constructs a division ring through skew-polynomial Ore fractions; the Laurent-series construction in Lemma~\ref{lem:embedding} is a direct alternative proof of the same embedding statement.

Both reset constructions have executable implementations with correctness proofs. The declarations \texttt{VerifiedAlgorithms.universal} and \texttt{VerifiedAlgorithms.sharp} also bound their work by fixed polynomials in the explicit input size. The cost model charges binary integer arithmetic, rational normalization, and finite list and table operations. Arithmetic bounds apply to the actual intermediate values and executed elimination stages; list and table operations receive specified conservative bulk charges. These are bounds in that model, not theorems about Lean's compiler or elapsed machine time. The definitions and scope are documented in \texttt{COST\_MODEL.md}.

The checked source uses Lean 4.28.0 and mathlib v4.28.0. The verification run builds all 97 source modules, checks the manuscript index against the compiled environment, runs executable checks, and audits the dependencies of all 2,970 declarations in the project namespace. The audit permits only the standard logical axioms \texttt{propext}, \texttt{Classical.choice}, and \texttt{Quot.sound}; there are no admitted proofs or additional axioms. From the formalization directory, the full check is
\begin{verbatim}
python3 checks/verify.py
\end{verbatim}
The complete Lean sources, verification report, source hashes, and axiom log are included as ancillary files with the arXiv submission.

OpenAI Codex assisted with the exposition, figures, and Lean formalization.

\end{document}